\documentclass[twocolumn]{myletters}
\group{Applied Integrable Systems}
\affiliation{National Institute of Technology, Gunma  College}{580, Toribamachi , Maebashi City, Gunma, 371-8530, Japan}
\affiliation{Department of Applied Mathematics, Waseda University}{3-4-1, Okubo, Shinjuku-ku, Tokyo 169-8555, Japan}
\authorinfo{Kazushige Endo}{1*}{11922960kz@gmail.com}
\authorinfo{Daisuke Takahashi}{2}{daisuket@waseda.jp}
\email{11922960kz@gmail.com}
\title{Three-dimensional fundamental diagram of particle system of 5 neighbors with two conserved densities}
\abstract{We discuss a particle system of 5 neighbors with two independent conserved densities.  The mean momentum uniquely depends on a pair of the densities and a three-dimensional fundamental diagram is obtained.  It shows the phase transition of behavior of asymptotic solution to the system.  Moreover, we propose two other systems which have the similar unique dependency obtained numerically.}
\usepackage{amsthm}
\usepackage{pdfpages}
\theoremstyle{definition}
\newtheorem{theorem}{Theorem}
\newtheorem{lemma}[theorem]{Lemma}
\begin{document}

\maketitle

\section{Introduction}
Fundamental diagram is an important object showing the asymptotic behavior of solutions to the system where particles move in the space sites\cite{fuks,nishinari}.  It shows the relation between the density of particles and their mean momentum.  There are interesting phenomena `phase transition' such that the momentum suddenly changes at some critical points of density.  For example, a phase transition from free flow to congested flow in a traffic system is well-known and has been studied theoretically using various mathematical models\cite{nagel,fukui,bando}.\par
  However, the density is not a unique parameter to determine the phase of particle flow.  There are many systems which give different mean momenta according to the variation of initial data for the same density\cite{nishinari2}.  To solve this redundancy, some other quantities should be necessary to determine the momentum in addition to the density.\par
  Our purpose of this letter is to show a three-dimensional framework to the fundamental diagram using two conserved densities\cite{loder}.  First, we propose an exact analysis on solutions to a particle system and derive a three-dimensional fundamental diagram.  The mean momentum is uniquely determined by a pair of particle density and another conserved density.  Second, we report two other relevant systems of which similar diagrams are obtained numerically.
\section{Particle system with two conserved quantities}
Let us consider the time evolution equation of a particle system as follows:
\begin{equation}  \label{evolv}
\begin{aligned}
  u_j^{n+1}=u_j^n&+q(u_{j-2}^n,u_{j-1}^n,u_j^n,u_{j+1}^n)\\
&-q(u_{j-1}^n,u_j^n,u_{j+1}^n,u_{j+2}^n),
\end{aligned}
\end{equation}
where $j$ is a site number, $n$ is an integer time, $u\in\{0,1\}$ is a binary state value and the flux $q(a,b,c,d)$ is given by the binary table shown in Table~\ref{tbl:flux}.
\begin{table}
\begin{center}
$
\renewcommand{\arraycolsep}{3pt}
\begin{array}{|c|c|c|c|c|c|c|c|}
\hline
 1111 & 1110 & 1101 & 1100 & 1011 & 1010 & 1001 & 1000 \\
\hline
  1 & 1 & 1 & 1 & 0 & 0 & 0 & 0 \\
\hline
\hline
 0111 & 0110 & 0101 & 0100 & 0011 & 0010 & 0001 & 0000 \\
\hline
  0 & 1 & 0 & 0 & 0 & 0 & 0 & 0 \\
\hline
\end{array}
$
\end{center}
\caption{Rule table of $q(a,b,c,d)$.  Upper and lower rows denote $(a,b,c,d)$ and $q(a,b,c,d)$ respectively.}
\label{tbl:flux}
\end{table}
If $a=b=1$ or $(a,b,c,d)=(0,1,1,0)$, $q(a,b,c,d)=1$ and otherwise 0.  Assume a periodic boundary condition for space sites with period $L$, that is, $u_{j+L}^n=u_j^n$.\par
  Consider that $u_j^n=1$ denotes a particle existing at site $j$ and time $n$ and $u_j^n=0$ an empty site.  Since \eqref{evolv} follows the conservation form, $\sum_{j=1}^Lu_j^n$ is constant for $n$.  Thus, the density $\rho=\sum_{j=1}^Lu_j^n/L$ is also.  Moreover, flux $q(u_{j-2}^n,u_{j-1}^n,u_j^n,u_{j+1}^n)$ means the number of particles moving from site $j-1$ to $j$ from $n$ to $n+1$.
\par
\begin{figure}[hbt]
\begin{tabular}{c}
\includegraphics[scale=0.5]{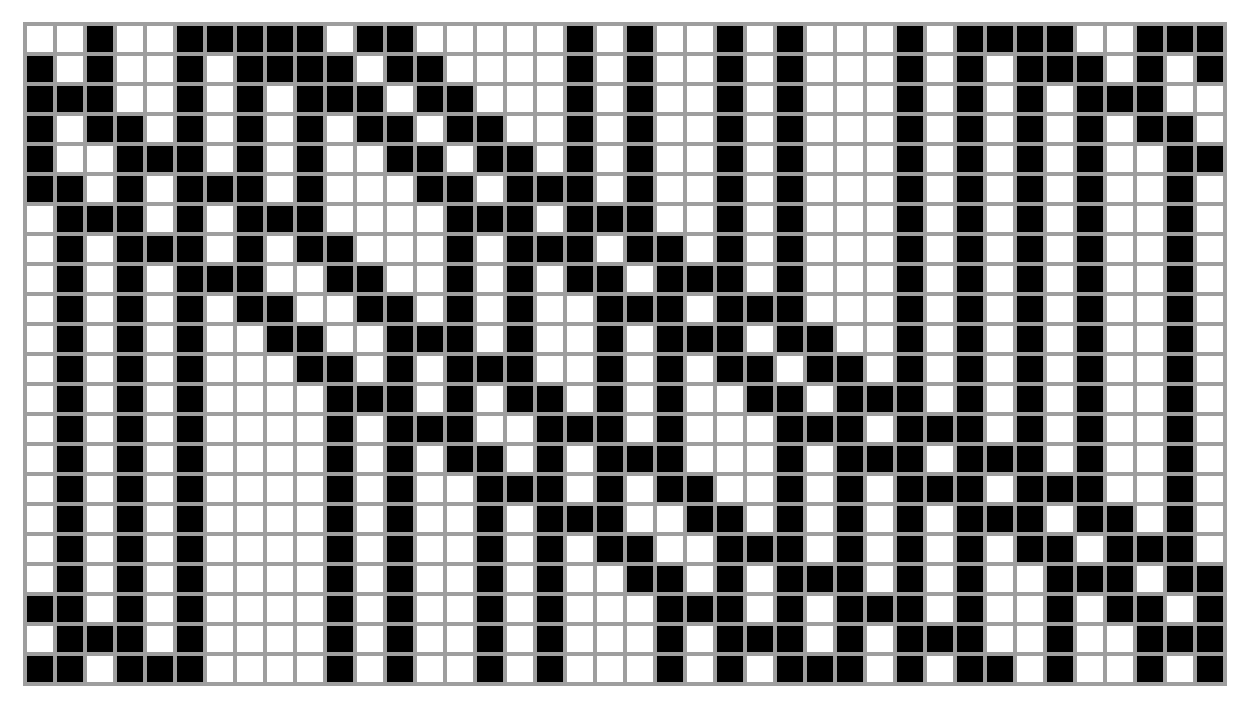} \\
(a) $\rho=0.5$\medskip\\
\includegraphics[scale=0.5]{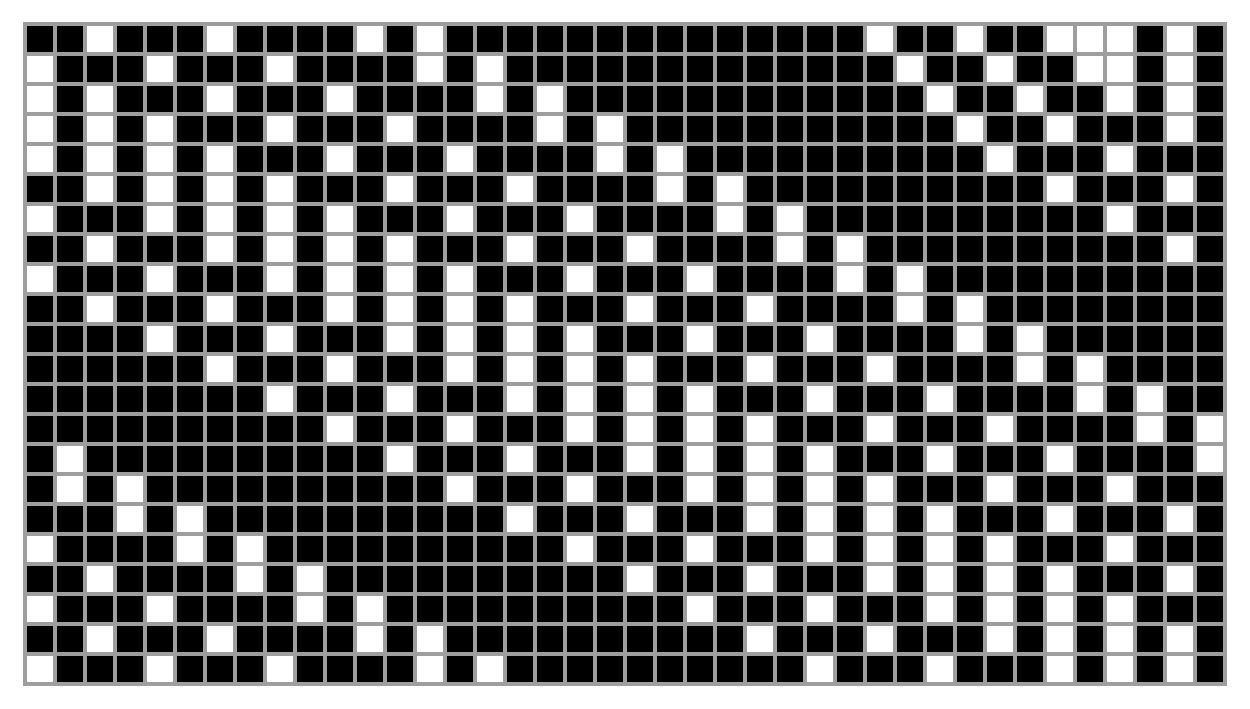}\\
(b) $\rho=0.75$
\end{tabular}
\caption{Examples of solutions to \eqref{evolv}.  Black square $\blacksquare$ and white square $\square$ denote $u=1$ and 0 respectively.  Initial data is shown at the top line and the state evolves downwards.}
\label{fig:evolv}
\end{figure}

  Equation \eqref{evolv} means the following motion rule of particles.
\begin{itemize}
\item
  An isolated particle (010) does not move.
\item
  For a pair of adjacent two particles (0110), both particles move.
\item
  For a sequence of more than two particles ($011\ldots10$), particles other than the leftmost move.
\end{itemize}
Figure~\ref{fig:motion} shows the motion of particles and the state of adjacent empty sites from time $n$ to $n+1$.  Note that the symbol \verb/*/ denotes an indefinite site of which value is dependent on values of other sites at time $n$.
\begin{figure}[hbt]
\begin{center}
$\begin{array}{lllllll}
n  &: & \verb/010/ & \verb/0110/ & \verb/01110/ & \verb/011110/ & ...\\
n+1&: & \verb/*10/ & \verb/*011/ & \verb/*1011/ & \verb/*10111/ & ...
\end{array}$
\end{center}
\caption{Motion of particles.}
\label{fig:motion}
\end{figure}
Figure~\ref{fig:evolv} shows two examples of solution to \eqref{evolv}.
\par
  There exists another conserved quantity for \eqref{evolv}.  Let us define $\#(x_1x_2\ldots x_k)^n$ by the number of local patterns $x_1x_2\ldots x_k$ included in the space sites at time $n$.  Then, $\#011^n$ is a conserved quantity since the pattern 011 at time $n+1$ are produced only from the patterns 0111 or 0110 as shown in Fig.~\ref{fig:011}.
\begin{figure}[hbt]
\begin{center}
$\begin{array}{llll}
n  &: & \verb/0111*/ & \verb/0110/\\
n+1&: & \verb/*1011/ & \verb/*011/
\end{array}$
\end{center}
\caption{Two patterns which produce {\tt 011} at the next time.}
\label{fig:011}
\end{figure}
Thus we obtain $\#011^{n+1}=\#0111^n+\#0110^n=\#011^n$.  Note that $\#1^n=\sum_j u_j^n$ is conserved and $\#0^n=K-\#1^n$ also.  Moreover, $\#110^n=\#011^n$ since $\#110^n+\#010^n=\#10^n$, $\#011^n+\#010^n=\#01^n$ and $\#01^n+\#11^n=\#10^n+\#11^n=\#1^n$ always hold.  Therefore, there are two independent conserved quantities $\#1^n$ (or $\#0^n$) and $\#011^n$ (or $\#110^n$) for \eqref{evolv}.  We omit their superscript $n$ as $\#1$ ($\#0$) and $\#011$ ($\#110$) to indicate their conservation.
\section{Asymptotic behavior}
  Let us introduce a notation $0k0$ ($k\in\mathbb{N}$) which means a local pattern of a sequence of 1's of length $k$ neighbored on both sides by 0.  For example, $020=0110$ and $\#030^n=\#01110^n$.  We obtain the following lemma on $\#0k0^n$ for $k\ge4$.
\begin{lemma}  \label{lemma:0k0}
  For large enough $n$ ($n\gg0$), $\#0k0^n$ ($k\ge4$) of any solution to \eqref{evolv} becomes constant for $n$.
\end{lemma}
The proof of this lemma is as follows. Every 1 moves to the right by one site or stays at the same position for each time.  Every stationary 1 is neighbored to the right by 0 at the next time from Fig.~\ref{fig:motion}.  Moreover, 1's other than the leftmost move to the right for the sequence $0k0$ ($k\ge4$).  Thus $\#0k0^n$ of maximum $k$ included in the state can not grow, decreases monotonically for $n$ and converges after enough time.  Then, $\#0k0^n$'s for smaller $k$'s follow the same process recursively and all $\#0k0^n$'s converge and become constant for $n\gg0$.  Thus, the lemma holds.  Note that the lemma does not hold for $k\le3$ since both 1's of 0110 move to the right.\par
  If all $\#0k0^n$'s for $k\ge4$ become constant for $n$, every sequence $0k0$ can be maintained by two types of evolutions shown in Fig.~\ref{fig:keep} and we can easily show the following lemma considering the stable evolution of $0k0$'s for $k\ge4$.
\begin{figure}[hbt]
\begin{center}
$\begin{array}{llll}
n  &: & \verb/0111...11010/ & \verb/0111...110111*/ \\
n+1&: & \verb/*1011...1110/ & \verb/*1011...111011/
\end{array}$
\end{center}
\caption{Two cases that $0k0$ keeps its length.}
\label{fig:keep}
\end{figure}
\begin{lemma}  \label{lemma:asymp}
  Assume that one or more local patterns 040, 050, $\ldots$ exist for $n\gg0$.  Then, they move to the right by two sites for every time.  All other 1's are included 010 or 030.  All 0's are isolated, that is, exist as 101.  Any $\#0k0^n$ ($k\in\mathbb{N}$) is constant for $n$ in this asymptotic solution.
\end{lemma}
  There is a special case of asymptotic solutions added to those described in Lemma~\ref{lemma:asymp}.  If all 1's are included in the local patterns 030 or 010 and all 0's are isolated, the state is an asymptotic solution and any $\#0k0^n$ is constant.\par
\begin{figure}[hbt]
\begin{center}
\includegraphics[scale=0.5]{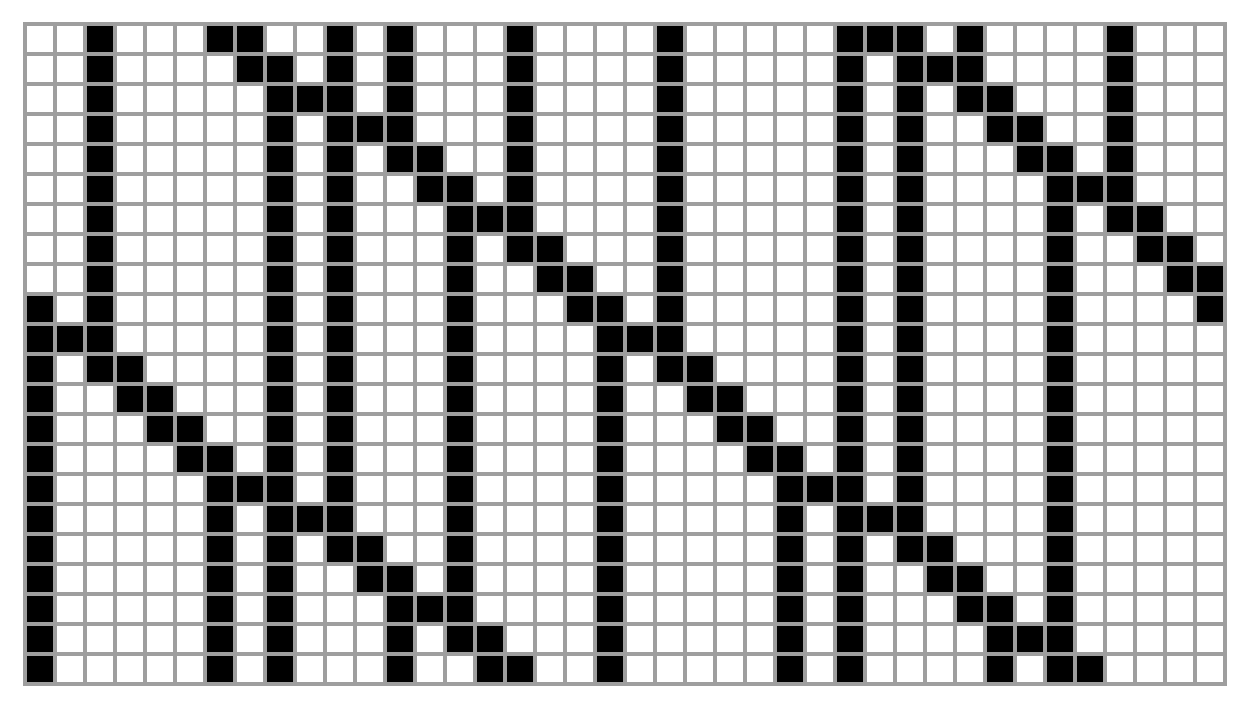}
\end{center}
\caption{Merging and splitting among local patterns 0110, 010 and 030.}
\label{fig:030}
\end{figure}

 To summarize asymptotic solutions of this special case and of Lemma~\ref{lemma:asymp}, they can be realized by the condition $\#0110^n=\#00^n=0$.  Therefore, $\#0110^n+\#00^n>0$ is necessary for another type of asymptotic solution.  Moreover, $\#0k0^n=0$ holds for any $k\ge4$ from Lemma~\ref{lemma:asymp} in the latter type of solution.  Since 0110 and 010 can merge and  produce 030 and conversely 030 can split into 0110 and 010, description of asymptotic behavior of this type of solution is more complicated than that of the former type.  Figure~\ref{fig:030} shows an example of solution where merging and splitting among 0110, 010 and 030 occur.
\par
However, the conditions about two types of asymptotic solutions shown above provide enough information to derive a three-dimensional fundamental diagram.  We give the following theorem to distinguish two types of asymptotic solutions as a consequence of this section.
\begin{theorem}  \label{theorem:type}
Asymptotic solution to \eqref{evolv} for $n\gg0$ obeys either of the following conditions.
\begin{itemize}
\item[(A)]
$\#0110^n=\#00^n=0$.
\item[(B)]
$\#0110^n+\#00^n>0$ and $\#0k0^n=0$ for any $k\ge4$.
\end{itemize}
\end{theorem}
\section{Three-dimensional fundamental diagram}
  First, we show that two types (A) and (B) described in Theorem~\ref{theorem:type} are distinguished by the signature of $2\#011-\#1+\#0$.  The following formulas always hold for any state.
\begin{equation*}
  \#1=\sum_{k\ge1}k\#0k0^n,\qquad \#0=\sum_{k\ge1}\#0k0^n+\#00^n.
\end{equation*}
For type (A), using these formulas and noting $\#0110^n=\#00^n=0$ we have
\begin{equation*}
\begin{aligned}
& 2\#011-\#1+\#0 \\
={}&
2\sum_{k\ge3}\#0k0^n-\Bigl(\#010^n+\sum_{k\ge3}k\#0k0^n\Bigr) \\
&\qquad +\Bigl(\#010^n+\sum_{k\ge3}\#0k0^n\Bigr) \\
={}&-\sum_{k\ge4}(k-3)\#0k0^n\le0.
\end{aligned}
\end{equation*}
For type (B), noting $\#0110^n+\#00^n>0$ and $\#0k0^n=0$ for any $k\ge4$, we have
\begin{equation*}
\begin{aligned}
& 2\#011-\#1+\#0 \\
={}&2(\#0110^n+\#030^n)-(\#010^n+2\#0110^n+3\#030^n) \\
&\qquad+(\#010^n+\#0110^n+\#030^n+\#00^n) \\
={}&\#0110^n+\#00^n>0
\end{aligned}
\end{equation*}
Therefore, the type (A) is in the case of $2\#011\le\#1-\#0$ and (B) $2\#011>\#1-\#0$.\par
  The mean momentum $Q$ is given by counting the number of moving particles as follows.
\begin{equation*}
  Q=\frac{1}{L}\Bigl(\sum_{k\ge3}(k-1)\#0k0^n+2\#0110^n\Bigr).
\end{equation*}
We have
\begin{equation*}
  Q=\frac{1}{L}\Bigl(\sum_{k\ge3}(k-1)\#0k0^n\Bigr)=\frac{1}{L}(\#1-\#0),
\end{equation*}
for type (A) and
\begin{equation*}
  Q=\frac{1}{L}(2(\#030^n+\#0110^n))=\frac{2}{L}\#011,
\end{equation*}
for type (B) considering the conditions for both types.  \par
\begin{figure}[hbt]
\begin{center}
\includegraphics[scale=0.25]{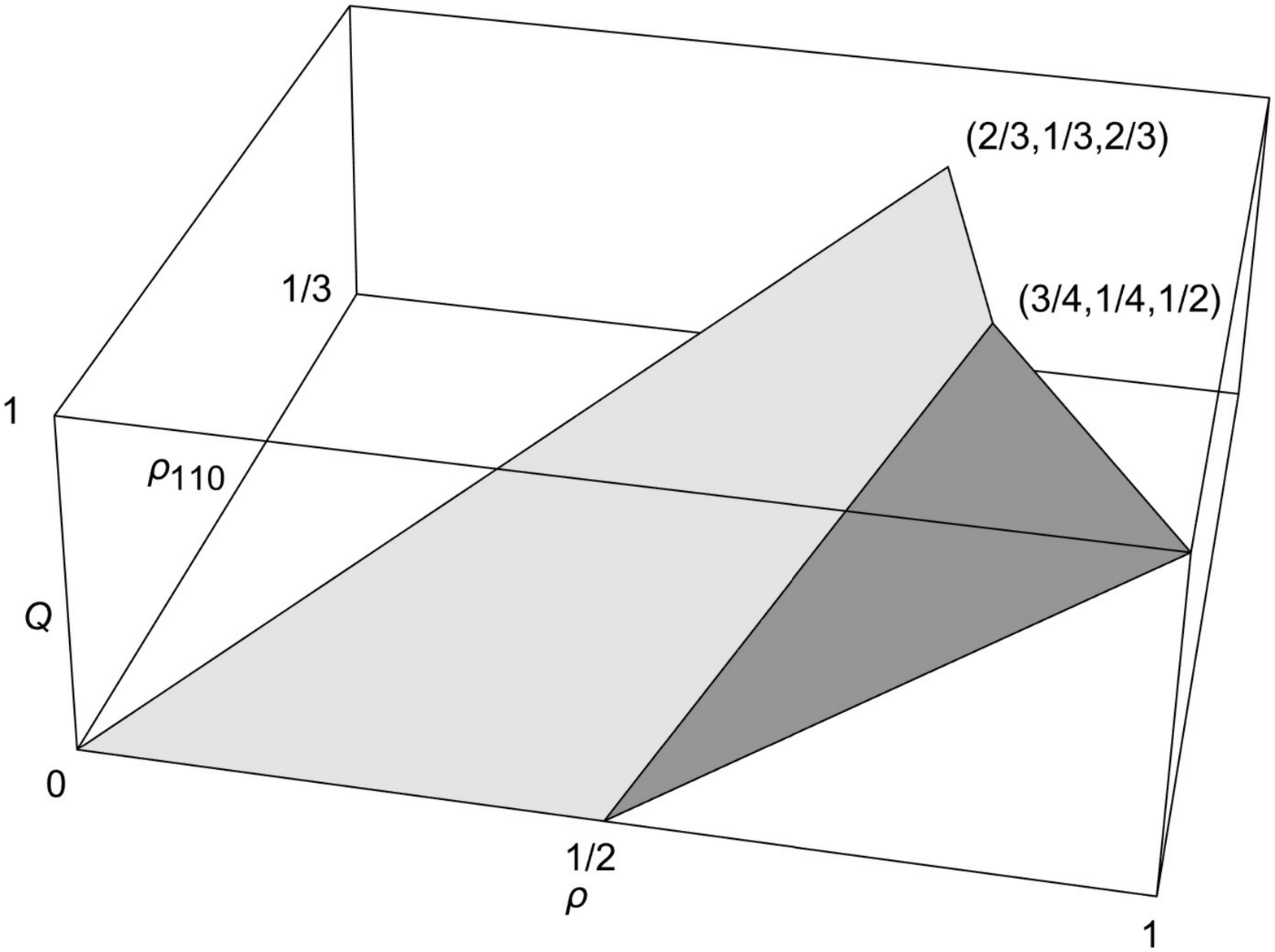}
\end{center}
\caption{Fundamental diagram of \eqref{evolv}.}
\label{fig:fd}
\end{figure}

  Finally, the following relation about the fundamental diagram is derived.
\begin{theorem}  \label{theorem:fd}
The mean momentum of \eqref{evolv} is uniquely given by two independent conserved densities $\rho=\#1/L$ and $\rho_{011}=\#011/L$ as follows.
\begin{equation*}
  Q=\max(\underbrace{2\rho-1}_{\text{\rm type (A)}},\,\underbrace{2\rho_{011}}_{\text{\rm type (B)}}).
\end{equation*}
\end{theorem}
Figure~\ref{fig:fd} shows the three-dimensional fundamental diagram obtained by Theorem~\ref{theorem:fd}.  There is a region of $\rho$ where both asymptotic solutions of type (A) and (B) appear from different initial data.  Therefore, another conserved density $\rho_{011}$ is a key density to distinguish them.

\section{Other systems}
There are many other particle systems in the form of \eqref{evolv} with $q$ different from that of Table~\ref{tbl:flux}.  We made numerical experiments on such systems and found some may have a three-dimensional fundamental diagram dependent on a conserved density or a quasi-conserved density in addition to the density $\rho$.  We show a rule table of $q(a,b,c,d)$ and a predicted $Q$ of two examples below.\par
Flux $q(a,b,c,d)$ of the first example is given by the following rule table.
\begin{equation*}
\renewcommand{\arraycolsep}{3pt}
\begin{array}{|c|c|c|c|c|c|c|c|}
\hline
 1111 & 1110 & 1101 & 1100 & 1011 & 1010 & 1001 & 1000 \\
\hline
  0 & 1 & 1 & 1 & 0 & 0 & 0 & 0 \\
\hline
\hline
 0111 & 0110 & 0101 & 0100 & 0011 & 0010 & 0001 & 0000 \\
\hline
  0 & 1 & 0 & 0 & 0 & 0 & 0 & 0 \\
\hline
\end{array}
\end{equation*}
From numerical results, we predict $Q=\min(2(1-\rho),\,\rho-\rho_{\rm odd})$ where $\rho_{\rm odd}$ is defined by $(\sum_{k=1}^\infty \#0(2k-1)0)/L$, that is, the number of sequences of 1's of odd length.  The density $\rho_{\rm odd}$ is conserved in this system since we can easily show any movable 1 is either of 1's included in a local pattern 110.\par
Flux $q(a,b,c,d)$ of the second example is given by the following rule table.
\begin{equation*}
\renewcommand{\arraycolsep}{3pt}
\begin{array}{|c|c|c|c|c|c|c|c|}
\hline
 1111 & 1110 & 1101 & 1100 & 1011 & 1010 & 1001 & 1000 \\
\hline
  0 & 1 & 1 & 1 & 0 & 1 & 0 & 1 \\
\hline
\hline
 0111 & 0110 & 0101 & 0100 & 0011 & 0010 & 0001 & 0000 \\
\hline
  0 & 0 & 1 & 1 & -1 & 0 & 0 & 0 \\
\hline
\end{array}
\end{equation*}
We predict $Q=\min(2\rho,\,-4\rho_{1*0}+1,\,2(1-\rho))$ where $\rho_{1*0}$ is the density of the number of local patterns 110 or 100.  Though the density $\rho_{1*0}$ is not a conserved density exactly, it becomes constant in the asymptotic solution for $n\gg0$ from any initial data considering the evolution rule.
\section{Concluding remarks}
  We made an exact analysis on a particle system of 5 neighbors in the conservation form using a flux.  The mean momentum is uniquely determined by two independent conserved densities and it provides a three-dimensional fundamental diagram.  Moreover, we proposed similar other systems using numerical experiments.\par
  One of future problems is to analyze them exactly and propose a common mechanism to give the three-dimensionality of fundamental diagram.  Moreover, there exist particle systems which have three or more conserved densities.  It is another future problem to find a higher dimensional dependency of mean momentum on conserved densities.\par
  Theoretical analysis on asymptotic behavior of particle systems can be applied to real transportation systems.  It is expected that information on mean momentum detailed by multiple conserved densities provides useful solution for transportation problems.
\references

\end{document}